\documentclass[preprint,3p,times]{elsarticle}
\usepackage{graphicx}
\usepackage{dcolumn}
\usepackage{bm}
\usepackage{bm}
\usepackage{mciteplus}
\usepackage{amsmath}
\usepackage{amsfonts,amssymb}
\usepackage{color}
\usepackage{cancel} 
\usepackage{enumerate,soul}
\usepackage{amssymb}
\usepackage[normalem]{ulem}

\begin{document}
\begin{frontmatter}

\title{On superstatistics and black hole quasinormal modes.}

\author{A. Mart\'inez-Merino$^a$}
\ead{aldo.martinez@academicos.udg.mx}
 \author{M. Sabido$^b$}%
\ead{msabido@fisica.ugto.mx}
\address{$^a$ Departamento de Ciencias Naturales y Exactas, CU Valles, Universidad de Guadalajara.\\ Carretera Guadalajara - 
 Ameca Km. 45.5,  C.P. 46600, Ameca, Jalisco, M\'exico.\\
  $^b$ Departamento  de F\'{\i}sica de la Universidad de Guanajuato,
 A.P. E-143, C.P. 37150, Le\'on, Guanajuato, M\'exico.}

\begin{abstract}
It is known that using Boltzmann-Gibbs statistics,  Bekenstein-Hawking entropy $S_{HB}$, and the quasinormal modes of black holes, one finds that the lowest value of  spin  is $j_{min}=1$. In this paper, we determine $j_{min}$, using non-extensive entropies that depend only on the probability (known as Obregon's entropies and have been derived from superstatistics). We also calculate $j_{min}$ for a set of non-extensive entropies that have free parameters and  are written in terms of $S_{BH}$. We find that $j_{min}$ depends on the area and the non-extensive parameter. 

For the non-extensive entropies that only depend on the probability, we find that the modification is only present for micro black holes. For classical black holes the results are the same as  for the Boltzmann-Gibbs statistics.

\end{abstract}
\end{frontmatter}



\section{Introduction}

Black holes are one of the most enigmatic and mysterious objects in physics. Recently,
direct verification of their existence was provided by the Event Horizon Telescope collaboration \cite{event}.
Despite the amount of research on the subject, there are several unanswered questions on black hole physics. 

Quantization of black holes was proposed in the pioneering work of Bekenstein \cite{bekenstein}. He suggested that the surface gravity is proportional to the temperature, and  the area of the event horizon is proportional to its entropy.  Moreover, he conjectured that the horizon area of non extremal black holes plays the role of a classic adiabatic invariant. Finally, he concluded that the horizon area should have a discrete spectrum with uniformly spaced eigenvalues,
\begin{equation}
A_n=\bar{\gamma}l_{p}^{2}n\label{ABeks},\qquad n=1,2,...
\end{equation}
where $\bar{\gamma}$ is a dimensionless constant.

With the development of Loop Quantum Gravity (LQG),  the correct spectrum of the area operator was obtained in \cite{ashtekar_1, krasnov}, being
\begin{equation}\label{spectrum}
A(j)=8\pi \gamma l^2_p\sqrt{j(j+1)},
\end{equation}
where $\gamma$ is the Immirzi parameter \cite{immirzi, rovelli}. It is a free parameter in Loop Quantum Gravity (LQG), and  determines the value of the minimal area.  As any fundamental constant in the theory we must find a way to determine its value, and this is where the entropy comes into play. The entropy is a quantity related to spectrum and therefore seems to be the main candidate to determine the value of $\gamma$.
In \cite{dreyer}, the author established a method to determine $\gamma$ using the quasinormal modes for the  Schwarzschild black hole. This approach relates the area (derived in LQG) with the mass and area of the Schwarzschild black hole. Using Boltzmann-Gibbs statistics, obtains the expression $\gamma=\frac{\ln{3}}{2\pi \sqrt{2}}$. 

In this work following \cite{dreyer}, we use quasinormal modes and Obregon's entropies \cite{obregon1}  to determine the minimum value $j_{min}$. These entropies are generalizations to the Boltzmann-Gibbs entropy, and only depend on the probability. Calculation of $j_{min}$ has been done for Tsallis entropy \cite{everton1}. The parameter $q$ is used to fix  $j_{min}=1/2$. Finally, we also study  no-extensive entropies that have free parameters and are explicitly a function of the Bekenstein-Hawking entropy. This approach was originally presented  in \cite{everton2} for Barrow's entropy.

The paper is organized as follows. In section \ref{sec_1},  we follow the approach {in} \cite{dreyer} for the non-extensive entropies that only depend on the probability. In section \ref{sec_2}, we workout the case for other non-extensive entropies that have free parameters and are written in terms of $S_{BH}$. Section \ref{final} is devoted discussion and final remarks.


\section{Black holes and non-extensive entropies that only depend on the probability.}\label{sec_1}

The loss of the extensive property is one of main complications that arise when generalizing  Boltzmann's entropy. Nonetheless, non-extensive entropies can be useful in the study of  more general phenomena. There is a large set of non-extensive entropies that are related in \textit{Superstatistics}. These entropies can be {derived using different temperature distributions} \cite{beck}. One of the most useful and better studied generalization  is  Tsallis entropy \cite{tsallis}. Tsallis\footnote{In terms of the probability, Tsallis entropy is given by $S_{q}=\frac{1}{1-q} \sum_{l=1}^{\Omega}p_l^q$.} entropy is non-extensive and has a free parameter $q$, known as the entropic index. The value of the entropic index is dependent on the physical system\footnote{Applications of Tsallis entropy are present in different areas of physics, ranging from high energy collisional experiments, velocity distributions in plasma, anomalous diffusion, quantum entangled systems, to name a few. } under study. For $q=1$, Tsallis entropy reduces to the usual Boltzmann-Gibbs entropy.

In the context of superstatistics, starting with a Gamma distribution for the temperature, in \cite{obregon1} the author derives  non-extensive entropies that depend only on the probability. These entropies are known as Obregon's entropies and in contrast to Tsallis entropy, don't have  free parameters. Furthermore, Obregon's entropies  in the limit of small probabilities (or equivalently, a large number of states), reduce  to Boltzmann-Gibbs statistics. It is worth mentioning that these entropies have been used in connection with entropic gravity \cite{pinki}, and  AdS/CFT \cite{obregon2}.

Let us explicitly give the functional form of Obregon's entropies. The first entropy is denoted by $\mathcal{S_+}$ and is given by the expression
\begin{equation}
\mathcal{S_+} = \sum_{l=1}^{\Omega} \left( 1 - p_l^{p_l}\right), \label{OnE}
\end{equation}
{with the probabilities satisfying the usual constraint} $\sum_l p_l = 1$.

There is also another entropy of the form 
\begin{equation}
\mathcal{S_-} = \sum_{l=1}^{\Omega} \left( p_l^{-p_l}-1\right), \label{two}
\end{equation}
and a third one defined from the sum of the previous entropies $S_\pm=\frac{1}{2}\left( S_+ + S_-\right)$. These three entropies are the only generalizations that depend only on the probability\footnote{In the expressions for these entropies we are dealing with  dimensionless quantities, thus we have already divided the entropy by the Boltzmann constant, $k_B$.}.

\subsection{Relating entropy to black hole quasinormal modes}
{Let us  briefly review the approach in \cite{dreyer}, to relate the entropy and quasinormal modes of black holes.} 
For a large imaginary part  the frequency of the quasinormal modes \cite{nollert} is
\begin{equation}\label{quasi}
M\omega_n=\frac{\ln{3}}{8\pi}+\frac{i}{4}\left ( n+\frac{1}{2}\right),
\end{equation}
the value of the real part of Eq.(\ref{quasi}) was previously proposed in \cite{hod}. The energy spectrum is
$\Delta M=l_p^2 w_n$ 
(where  we are using units $G=c=1$) and we have defined $w_n=\frac{1}{M}\mathcal{R}e(\omega_n)=\frac{\ln{3}}{8\pi M}$. 
Considering that $A=16\pi M^2$, when we introduce a change in the mass $\Delta M$ we get a change in the area $\Delta A=4 l_p^2 \ln{3}$. Using  Eq.(\ref{spectrum}),we find $\gamma$ as a function of $j_{min}$
\begin{equation}\label{immirzi}
\gamma=\frac{\ln{3}}{2\pi \sqrt{j_{min}(j_{min}+1)}}
\end{equation}

\noindent The number of microstates of the configurations in a punctured sphere is
\begin{equation}\label{omega}
\Omega = \prod_{n=1}^N (2 j_n + 1),
\end{equation}
where $j_n$ is the spin label of such punctures, and $N$ are the number of the punctures. Because the most important contributions come from configurations that satisfy  $j_n=j_{min}$, Eq.(\ref{omega}) takes the simpler form
\begin{equation}
\Omega = (2 j_{min} + 1)^N.
\end{equation}
The number of punctures $N$ is given by the ratio $A/\Delta A$,
\begin{equation}
N = \frac{A}{4l_p^2 \ln 3}.
\end{equation}
Now we can write $\Omega$ in terms of $j_{min}$ and the area of the event horizon,
\begin{equation}\label{omega2}
\Omega = (2 j_{min} + 1)^{A/ 4l_p^2 \ln 3}.
\end{equation}
Assuming a large  number of states  $\Omega$ ,
{from  Shannon's entropy 
and considering  equipartition, we get 
$\mathcal{S} = \ln \Omega.$ Moreover, using Eq.(\ref{omega}) and equating to the Bekenstein-Hawking entropy one finds that $j_{min}=1$.
Finally, from Eq.(\ref{immirzi}) we can see that the Immirzi parameter is  $\gamma = \frac{\ln 3}{2 \pi \sqrt{2}}$.}

As  we are interested in determining  $j_{min}$ for the non-extensive entropies that only depend on the probability, we simply follow the same approach, but instead of using Shannon's entropy we use the entropy in Eq.(\ref{OnE}).  

Working in the microcanonical ensemble and assuming equipartition, the probability of finding the system in a particular state is
equal to the inverse of the number of states, $p=p_l=\Omega^{-1}$. Then Eq.(\ref{OnE}) takes the form
\begin{subequations}\label{obregons}
\begin{equation}
\mathcal{S_+} = \sum_{l=1}^{\Omega} \left( 1 - \left(\frac{1}{\Omega}\right)^\frac{1}{\Omega} \right) = \Omega \left( 1 - 
\left(\frac{1}{\Omega}\right)^\frac{1}{\Omega} \right).\label{obregonsa}
\end{equation}
{With the same assumptions, we also rewrite $\mathcal{S_-}$ and $\mathcal{S_\pm}$}
\begin{eqnarray}
\mathcal{S_-} &=& \sum_{l=1}^{\Omega} \left( \left(\frac{1}{\Omega}\right)^{-\frac{1}{\Omega}} -1\right) = \Omega \left( \left(\frac{1}{\Omega}\right)^{-\frac{1}{\Omega}} -1\right),\label{obregonsb}\\
\mathcal{S_\pm} &=& \frac{1}{2}\sum_{l=1}^{\Omega}  \left(\left(\frac{1}{\Omega}\right)^{-\frac{1}{\Omega}}-\left(\frac{1}{\Omega}\right)^\frac{1}{\Omega}\right)= \Omega \left(\left(\frac{1}{\Omega}\right)^{-\frac{1}{\Omega}}-\left(\frac{1}{\Omega}\right)^\frac{1}{\Omega}\right).\label{obregonsc}
\end{eqnarray}
\end{subequations}
To determine $j_{min}$ we must find $\Omega$ as a function of $A$. For this we equate the Bekenstein-Hawking entropy to the non-extensive  entropies 
\begin{subequations}\label{jmas}
\begin{eqnarray}
p \frac{A}{4 l_p^2} &=& 1 - p^p,\,\qquad \textrm{for}\quad\mathcal{S_+},\label{jmasa}\\
p \frac{A}{4 l_p^2} &=& p^{-p}-1,\qquad \textrm{for}\quad\mathcal{S_-},\label{jmasb}\\
p \frac{A}{4 l_p^2} &=& \frac{1}{2}\left(p^{-p} - p^p\right ),\,\textrm{for}\quad \mathcal{S_\pm}.\label{jmasc}
\end{eqnarray}
\end{subequations}
Solving for $p$ gives the solution  for $\Omega$ and  after substituting in Eq.(\ref{omega2}) we obtain $j_{min}$. For  
large $\Omega$ we can do a series expansion for small $p$.
To first order in $p$, one simply reproduces the  result derived from Shannon's entropy. Deviations for $j_{min}$ are obtained by considering higher order terms in the expansion
\begin{equation}
p \frac{A}{4 l_p^2} = - p \ln p - \frac{1}{2} p^2 \ln^2 p - \frac{1}{6} p^3 \ln^3 p + \dots., \label{Eqa1}
\end{equation}
this is the expansion for $\mathcal{S_+}$. As expected similar expressions are derived for $\mathcal{S_-}$ and $\mathcal{S_\pm}$.
Using the solution for $p$ in Eq.(\ref{omega}) we finally find $j_{min}$ as a function of the area
\begin{eqnarray}
&&2 j_{min} + 1 =\\
& & \exp \ln 3 \left\{ 1 + \frac{1}{2} \frac{A}{4 l_p^2} e^{-\frac{A}{4 l_p^2}} + \frac{1}{12} \left( \frac{A}{4 l_p^2} \right)^2
\left(4 - 3 \frac{A}{4 l_p^2} \right) e^{- \frac{2A}{4 l_p^2}}  + \frac{1}{48} \left( \frac{A}{4 l_p^2} \right)^3 \left( 10 -24 \frac{A}{4 l_p^2} + 9 \left(\frac{A}{4 l_p^2}\right)^2 \right) 
e^{-\frac{3A}{4 l_p^2}} + \dots \right\}.\nonumber
\end{eqnarray}
We can see that higher order terms are exponentially suppressed and we have a good approximation using the first three terms.
\begin{figure}[htbp] 
   \centering
   \includegraphics[width=5in]{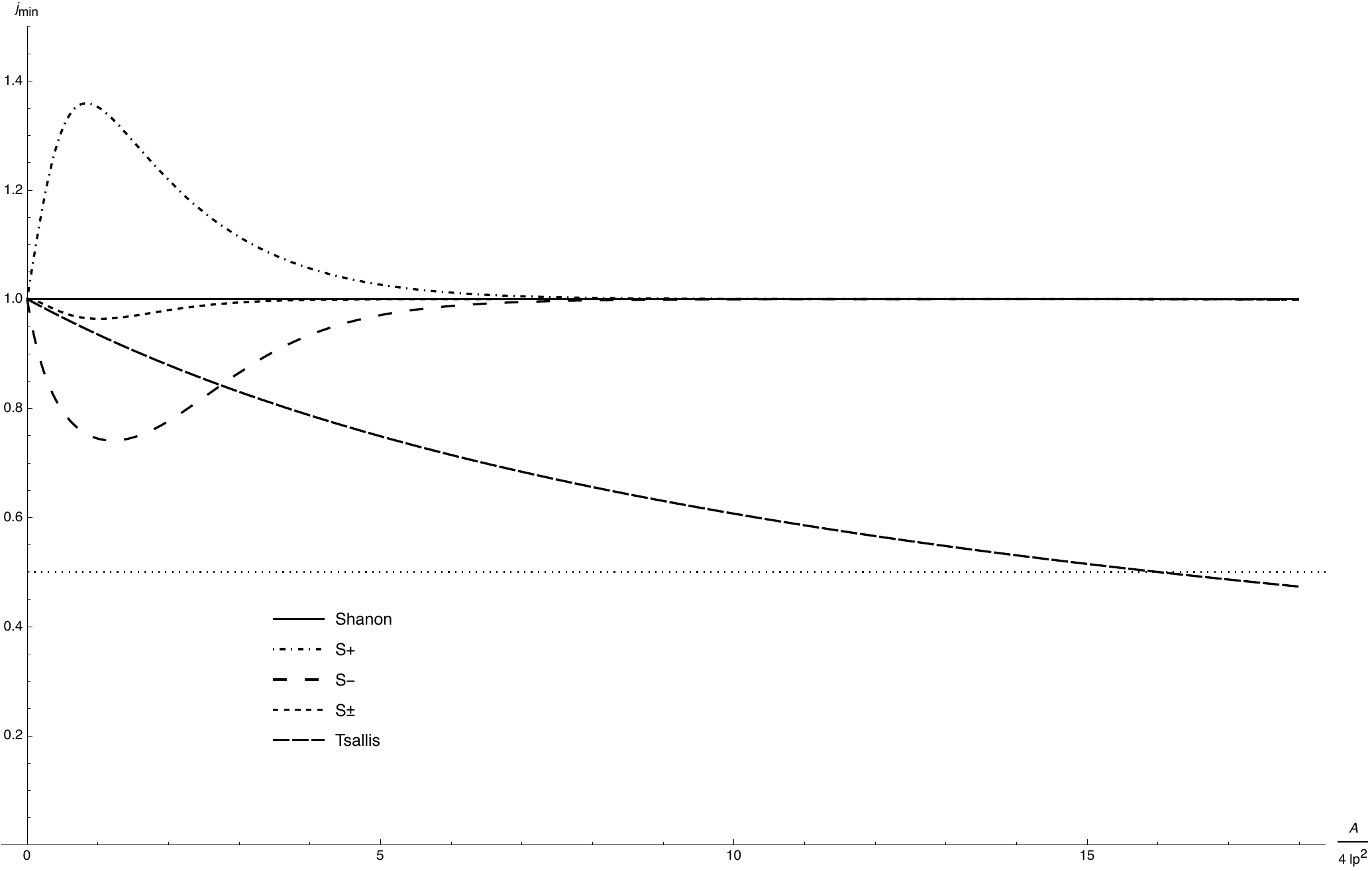} 
   \caption{{We have plotted $j_{min}$ for the different non-extensive entropies that only depend on the probability and for  Tsallis entropy.  We can see that for Planck scale black holes $j_{min}\ne 1$, but for large ones, the result from Shannon's entropy is recovered. For Tsallis entropy it is possible to find a value of $q$ that for a particular area $j_{min}=1/2$}. For arbitrary large area with a fixed value of $q$, $j_{min}$ goes to zero. }
   \label{figura1}
\end{figure}
Following the same steps as before, we can calculate $j_{min}$ using the entropy $\mathcal{S_{-}}$. From Eq.(\ref{jmasb}) we obtain the area as a function of $p$
\begin{equation}
p \frac{A}{4 l_p^2} = - p \ln p + \frac{1}{2} p^2 \ln^2 p - \frac{1}{6} p^3 \ln^3 p + \dots,
\end{equation}
solving for $p$, we arrive at the equation for $j_{min}$ for $\mathcal{S_{-}}$
\begin{eqnarray}
& &2 j_{min} + 1 =\\ 
& &\exp \ln 3 \left\{ 1 - \frac{1}{2} \frac{A}{4 l_p^2} e^{-\frac{A}{4 l_p^2}} + \frac{1}{12} \left(\frac{A}{4 l_p^2}\right)^2
\left( 4 - 3\frac{A}{4 l_p^2} \right) e^{-\frac{2A}{4 l_p^2}}  - \frac{1}{48} \left( \frac{A}{4 l_p^2} \right)^3 \left( 10 - 24 \frac{A}{4 l_p^2} + 9 \left( \frac{A}{4 l_p^2} \right)^2 \right)
e^{-\frac{3A}{4 l_p^2}} + \dots \right\}.\nonumber
\end{eqnarray}
Finally, for  $\mathcal{S}_{\pm}$
we have for $j_{min}$ 
\begin{eqnarray}
& & 2 j_{min} + 1 = \exp \ln 3 \left\{ 1 - \frac{1}{6} \left( \frac{A}{4 l_p^2} \right)^2 e^{-\frac{2 A}{4 l_p^2}} \right. \\
& & + \frac{1}{360} \left( \frac{A}{4 l_p^2} \right)^4 \left( 27 - 20 \frac{A}{4 l_p^2} \right) e^{-\frac{4 A}{4 l_p^2}}  -  \left. \frac{1}{64800} \left( \frac{A}{4 l_p^2} \right)^6 \left( 2880 - 4860 \frac{A}{4 l_p^2} + 1800 \left( \frac{A}{4 l_p^2} \right)^2
\right) e^{-\frac{6 A}{4 l_p^2}} - \dots \right\}. \nonumber
\end{eqnarray}

For  $\mathcal{S}_+$, $\mathcal{S}_-$ and $\mathcal{S}_\pm$, when we consider the case of large area $A$ we obtain $j_{min}=1$.
Since these entropies only depend on the probability, there are no free parameters we can use to fix the value of $j_{min}$. 
We can see in Fig.(\ref{figura1}), that for small $A$ (micro black holes) we have a different value for $j_{min}$. Even though  $j_{min}<1$  for $\mathcal{S}_-$ and $\mathcal{S}_\pm$, we don't have an area for which $j_{min}=1/2$.
This in contrast to the result fromTsallis entropy, where 
$j_{min}$ is given by
\begin{equation}
2 j_{min}+1= \left[(1+(1-q)\frac{A}{4lp^2} \right]^\frac{4lp^2\ln{3}}{A(1-q)},
\end{equation}
the parameter\footnote{The derivation and analysis is done in reference \cite{everton1}. They find the range of values for $q$  to satisfy $j_{min}=1/2$.} $q$ can be used to change the value of $j_{min}$. The value for $q$ is constrained to  $-\frac{4lp^2}{A}<(1-q)<1.37\frac{4lp^2}{A}$. Although you can have $j_{min}=1/2$, you have  a different values of $q$ for different black holes. For Obregon's entropies, $j_{min}<1$ for quantum black holes, and for classical black holes one recovers the prediction of Boltzmann-Gibbs statistics.

\section{Non-extensive entropies with free parameters}\label{sec_2}
The entropies studied in the previous section seem to be the only generalization to Shannon's entropy that only depend on the probability. Nonetheless, there are several other generalizations that have free parameters {besides Tsallis entropy}. We will focus our attention to a set of entropies that have been used in connection to the black hole area entropy law. The expressions for these entropies are written in terms of the Bekenstein-Hawking entropy $S_{BH}$, therefore we will follow \cite{everton2} to simplify the derivation of $j_{min}$.

\subsection{Tsallis-Cirto entropy}

The Tsallis-Cirto entropy was proposed in \cite{tsalliscirto} to solve thermodynamic inconsistencies for the Schwarzschild black hole. This entropy is defined by the relation

\begin{equation}\label{tsallis}
S_{TC} = (S_{BH})^\delta,
\end{equation}
in the limit  $\delta\to 1$, the Bekenstein-Hawking entropy is recovered. Since the entropy in Eq.(\ref{tsallis}) is written in terms of $S_{BH}$, we equate to the logarithm of the number of states (the number of states is given in Eq.(\ref{omega2})), to obtain $j_{min}$ 
\begin{equation}\label{jmintsallis}
j_{min} = \frac{1}{2} \left[ 3^{\left( \frac{A}{4 l_p^2} \right)^{\delta-1}}-1\right].
\end{equation}
\begin{figure}[htbp] 
   \centering
   \includegraphics[width=5in]{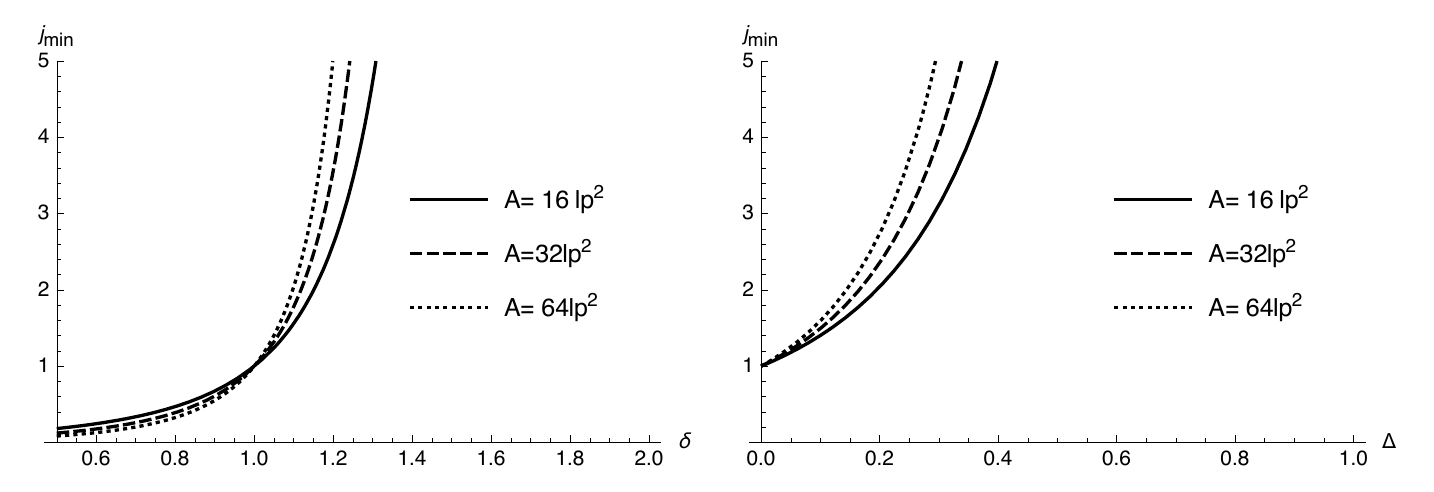} 
   \caption{{The first plot is the predicted $j_{min}$  for Tsallis-Cirto entropy. We can see that for a particular value of $A$ there is a corresponding value of $\delta$ that gives $j_{min}=1/2$. The second plot corresponds to Barrow's entropy. For $\Delta>0$, we can find an equivalent description in terms of Tsallis-Cirto entropy.}}
   \label{figura2}
\end{figure}
\newpage
For the Tsallis-Cirto entropy, we can satisfy $j_{min}=1/2$, for particular values of $\delta$ and the black hole horizon area 
\begin{equation}
\delta = \frac{\ln \frac{\ln 2}{\ln 3}}{\ln \frac{A}{4 l_p^2}}+ 1.
\end{equation}
{In Fig.(\ref{figura2}) we show $j_{min}$ as a function of $\delta$ for  values of the area. We can see  that it is possible to have $j_{min}= 1/2$.}
If we take $\delta=1+\frac{\Delta}{2}$ in Eq.(\ref{tsallis}) and Eq.(\ref{jmintsallis}), we obtain the results  to Barrow's entropy  $S_{Barrow}=\left(\frac{A}{4lp^2}\right)^{1+\Delta/2}$.
This entropy is related to a fractal structure on the black hole surface, is of quantum origin and is encoded in the parameter $\Delta$. When using the Barrow's interpretation, one gets $j_{min}>1$ for $\Delta> 0$. 


\subsection{Modified R\'enyi entropy}

We now consider the {modified R\'enyi entropy}. In connection to the black hole area entropy law, this entropy is  given by the relation
\begin{equation}
S_{MR} = \frac{1}{\lambda} \ln (1 + \lambda S_{BH}),
\end{equation}
where $\lambda$ is a positive constant. In the limit $\lambda \rightarrow 0$ we recover the Bekenstein-Hawking entropy. {As in the previous case, this entropy is written in terms of $S_{BH}$. Therefore we equate to the logarithm of the number of states of the black hole. 
Using the expression for the Bekenstein-Hawking entropy,} the result for $j_{min}$ is
\begin{equation}
j_{min} = \frac{1}{2} \left[ 3^{f(A; \lambda)} - 1\right],
\end{equation}
where

\begin{equation}
f(A; \lambda) = \frac{4 l_p^2}{\lambda A} \ln\left( 1+ \lambda\frac{A}{4 l_p^2} \right).
\end{equation}
It is easy to verify that in the limit when $\lambda\to 0$, $j_{min}\to 1$, recovering the result for the Bekenstein-Hawking entropy.
\begin{figure}[htbp] 
   \centering
   \includegraphics[width=4.5in]{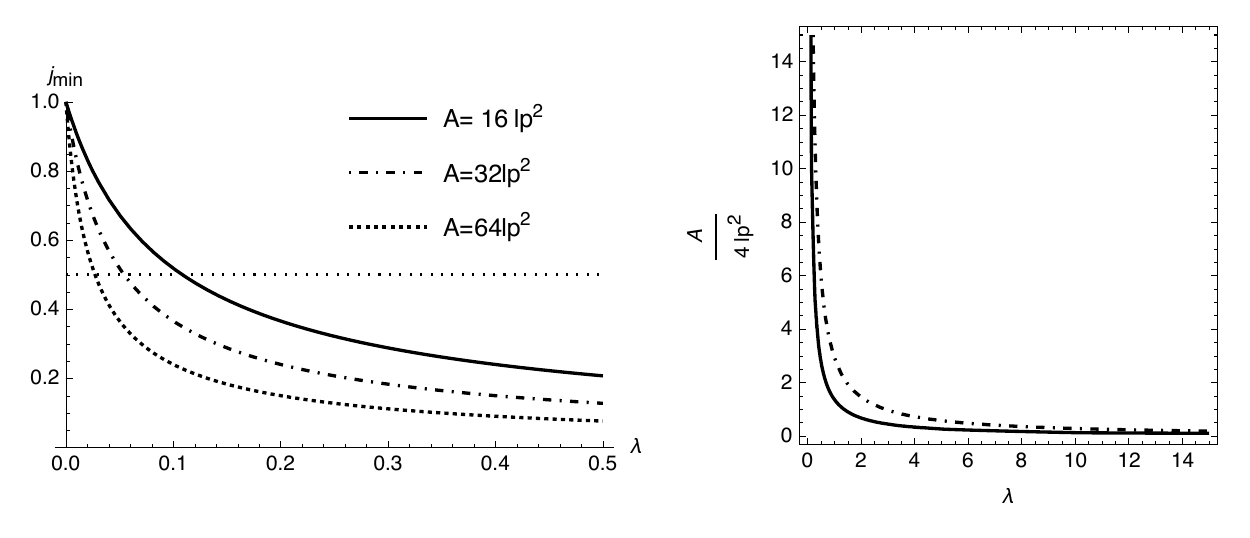} 
   \caption{The first graph is a plot of  $j_{min}$ derived from the modified R\'enyi entropy. We can see in this plot that to have $j_{min}=1/2$ the value of $\lambda$  is negative. Therefore, for this entropy  $j_{min}\ge 1$. The second plot is for the values of $\lambda$ and $A$ that give $j_{min}=1/2$. The solid line corresponds to the numerical solution to Eq.(\ref{renyi_eq}), the dotted line for the approximation.}
   \label{figura3}
\end{figure}
Looking for the value of $\lambda$ which gives us $j_{min} = 1/2$, we arrive at the next transcendental equation
\begin{equation}\label{renyi_eq}
\ln \left( 1 + \frac{\lambda A}{4 l_p^2} \right) = \frac{\ln 2}{\ln 3} \frac{\lambda A}{4 l_p^2}.
\end{equation}
The case $\lambda = 0$, although a solution is excluded since for this value the entropy goes to Bekenstein-Hawking $j_{min} = 1$. For large $\lambda$ we find the approximate relation  $\lambda=\frac{8 lp^2}{3 A \ln[3]}$.


\subsection{Sharma-Mittal entropy}

The last entropy we consider is the Sharma-Mittal. This entropy was proposed to construct a new model for holographic dark energy \cite{jawad}. Moreover, it is a generalization of both the R\'enyi and Tsallis entropy. In connection to the black hole area entropy law, is defined by the relation
\begin{equation}
S_{SM} = \frac{1}{R} \left[ \left( 1 + \delta S_{BH} \right)^{R/\delta} - 1 \right].
\end{equation}
The entropy interpolates between the modified R\'enyi entropy ($R \rightarrow 0$) and Bekenstein-Hawking entropy ($R \rightarrow \delta$). {As in previous cases, we assume that $S_{SM} = \ln \Omega$}. Thus,
the value for $j_{min}$ is
\begin{equation}
j_{min} = \frac{1}{2} \left[ 3^{g(A; R, \delta)} - 1 \right],
\end{equation}
{where
\begin{equation}
g(A; R, \delta) = \frac{4 l_p^2}{R A} \left[ \left( \frac{\delta A}{4 l_p^2} + 1 \right)^{R/\delta} - 1 \right].
\end{equation}}

The definition of this entropy in terms of these two parameters, allows for the possibility of finding a region where the value of $j_{min}$ be equal to 1/2. In the next plot, we show that for certain values of the area which pairs of such values gives $j_{min} = 1/2$.
\begin{figure}[htbp] 
   \centering
   \includegraphics[width=3.5in]{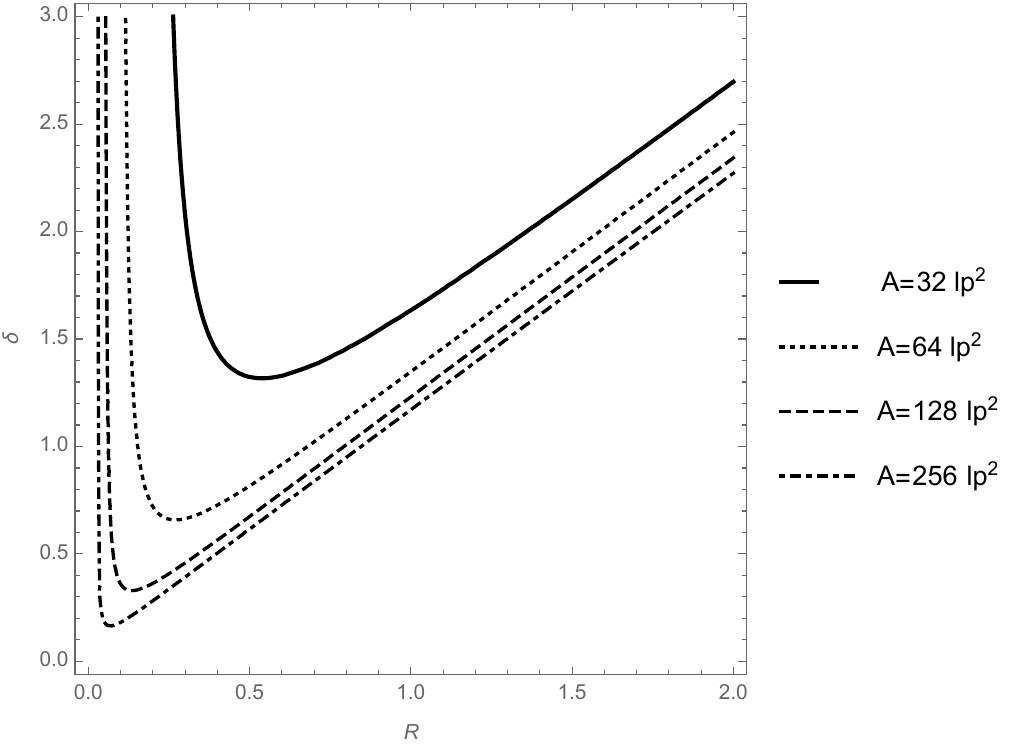} 
   \caption{Plot for the values of $R$ and $\delta$ in order to have $j_{min}$ for the the Sharma-Mittal entropy.}
   \label{figura4}
\end{figure}

\section{Final remarks}\label{final}
In this paper we have used quasinormal modes  to determine the minimum value $j_{min}$for non-extensive entropies. These entropies are  generalizations  of  the Boltzmann-Gibbs entropy. 
We worked out two classes of non-extensive entropies, the first class only depend on the probability. The second class  have free parameters and are written in terms of Bekenstein-Hawking entropy.

With respect to the non-extensive entropies that only depend on the probability, the only entropies that satisfy this requirement are the so called Obregon's entropies. Assuming equipartition we find $j_{min}$ for these entropies.
For $S_{-}$ and $S_\pm$ we see from Fig.(\ref{figura1}) that the minimum value is less than one, but for large area we recover the usual value. Therefore, we can conclude that the modification from using these entropies is only present for micro black holes. This is consistent with results obtained for fluids \cite{torres}, where it is showed that the effects of these entropies are not present for classical systems. Moreover,  for $A>8l_p^2$ the result is the same as BG, therefore we conclude that that the effects of using non extensive entropies are only present on micro black holes.

Of the non-extensive entropies that have free parameters and are function of $S_{BH}$. The  free parameters on these entropies, allows us to fix $j_{min}=1/2$, for particular values of the $A$. Moreover, for specific values of their respective parameters, they  reproduce  the Boltzmann-Gibbs results. Using this class of non-extensive entropies $SO(3)$ and $SU(2)$ spin networks are valid. It is worth mentioning that we can have  $j_{min}>1$ and therefore these entropies generalize the value of $j_{min}$.
\begin{figure}[htbp] 
   \centering
   \includegraphics[width=4.5in]{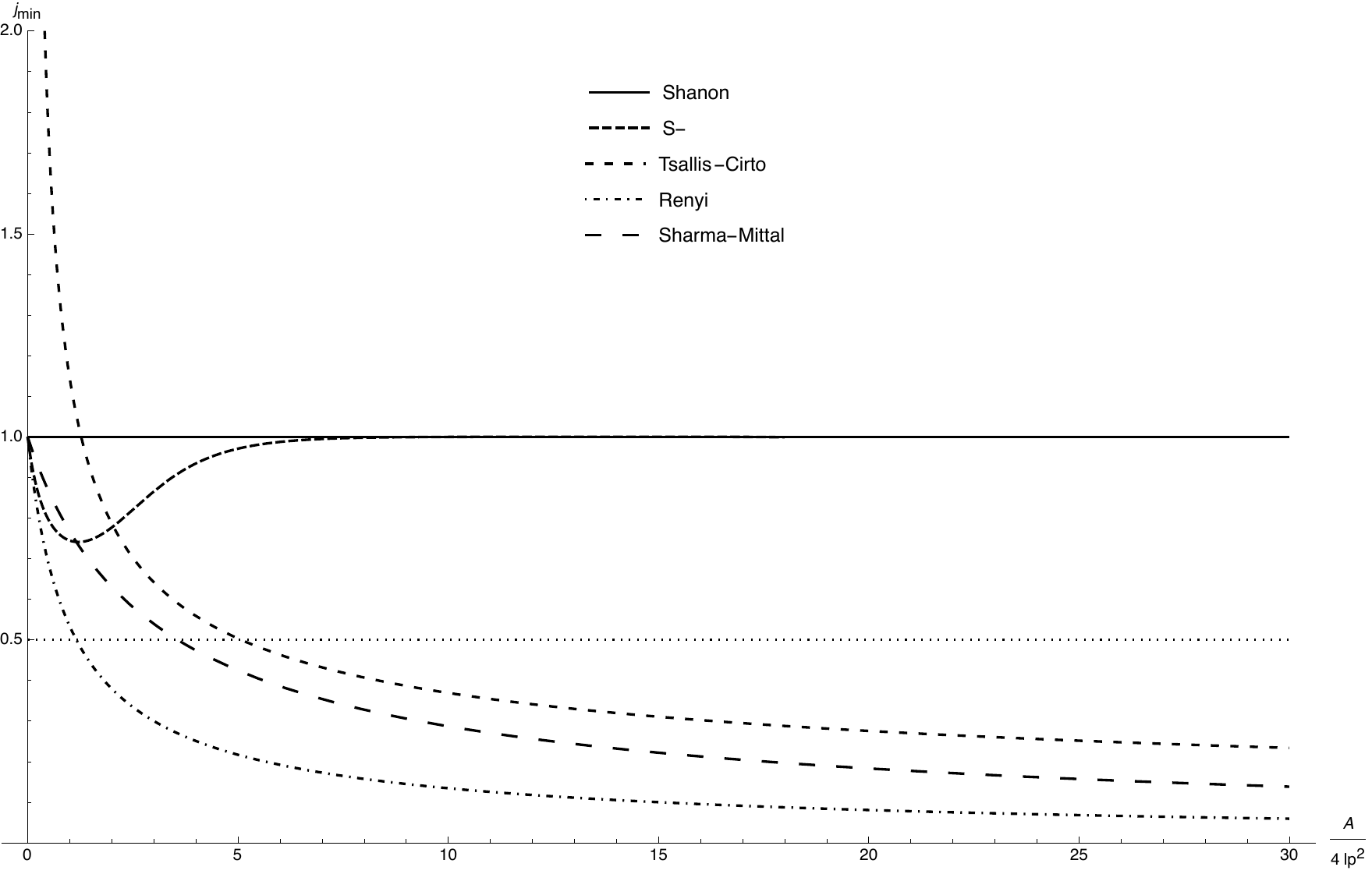} 
   \caption{Plot of $j_{min}$ as a function of the black hole area, derived from non-extensive entropies. }
   \label{figura5}
\end{figure}

{In summary, the use of non-extensive entropies modify $j_{min}$. For Obregon's entropies, $j_{min}\ne 1$  for micro black holes. Therefore, there is a  possibility that non-extensive statistics modifies the dynamics of quantum gravity. Of particular interest are the effects on cosmology, consequently the effects of no-extensive entropies could change the dynamics of the very early Universe. This is work in progress and will be reported elsewhere.}


\section*{Acknowledgements}
This work is supported by CONACYT grant 258982. M. S. is supported by CIIC-032/2021. A. M-M. would like to thank O. Obreg\'on for fruitful discussions, and acknowledges that this work began when he was part of C\'atedras Conacyt program.


\section*{Appendix}
To derive  $j_{min}$ as a function of the  black hole area, we must solve for $p$ in terms $A$. We begin by rewriting  Eq. (\ref{jmasa})
\begin{equation}
\frac{A}{4l_p^2} = \frac{1}{p} (1-p^p) = - \ln p - \sum_{n=2}^{\infty} p^{n-1} \ln^n p,
\end{equation}
or equivalently 
\begin{equation}
\ln p = -\frac{A}{4l_p^2} - \sum_{n=2}^{\infty} p^{n-1} \ln^n p = -\frac{A}{4l_p^2} - \sum_{n=2}^{\infty} \ln^n p e^{(n-1)\ln p}.
\end{equation}
This equation is of the form $y = x + f(y)$, with $y = \ln p$ and $x = -\frac{A}{4l_p^2}$ also $f (y) = 0$ when $y=0$.
Applying  Lagrange's inversion theorem \cite{Gross}, we can find a series solution for $y$ in terms of $x$, using the formula
\begin{equation}
y = x + \sum_{n=1}^{\infty} \frac{1}{n!} \left( \frac{d}{d x} \right)^{n-1} \{ f(x)^n \}.
\end{equation}
Using $n=3$ in the formula, the result is
\begin{eqnarray}
\ln p = - \frac{A}{4 l_p^2} - \frac{1}{2} \left( \frac{A}{4 l_p^2} \right)^2 e^{-\frac{A}{4 l_p^2}} - \frac{1}{12} \left( \frac{A}{4 l_p^2} \right)^3
\left(4 - 3 \frac{A}{4 l_p^2} \right) e^{- \frac{2A}{4 l_p^2}} - \frac{1}{48} \left( \frac{A}{4 l_p^2} \right)^4 \left( 10 -24 \frac{A}{4 l_p^2} + 9 \left(\frac{A}{4 l_p^2}\right)^2 \right) 
e^{-\frac{3A}{4 l_p^2}} - \dots,
\end{eqnarray}
the terms become exponentially suppressed for higher order terms, therefore we good accuracy for $n=3$.

We apply the same method for the entropies (\ref{obregonsb}) and (\ref{obregonsc}).

\bibliographystyle{apsrev4-1}

\end{document}